# Tunable electrochemistry with moiré flat bands and topological defects at twisted bilayer graphene

Yun Yu[1], Kaidi Zhang[1], Holden Parks[2], Mohammad Babar[2], Stephen Carr[3], Isaac M. Craig[1], Madeline Van Winkle[1], Artur Lyssenko[1], Takashi Taniguchi[4], Kenji Watanabe[5], Venkatasubramanian Viswanathan[2,6], D. Kwabena Bediako[1,7]*

[1] Department of Chemistry, University of California, Berkeley, CA 94720, USA.
[2] Department of Mechanical Engineering, Carnegie Mellon University, Pittsburgh, 15213, Pennsylvania, USA.
[3] Brown Theoretical Physics Center, Brown University, Providence, Rhode Island 02912, USA.
[4] International Center for Materials Nanoarchitectonics, National Institute for Materials Science, 1-1 Namiki, Tsukuba 305-0044, Japan
[5] Research Center for Functional Materials, National Institute for Materials Science, 1-1 Namiki, Tsukuba 305-0044, Japan
[6] Department of Physics, Carnegie Mellon University, Pittsburgh, 15213, Pennsylvania, USA.
[7] Chemical Sciences Division, Lawrence Berkeley National Laboratory, Berkeley, CA 94720, USA

*Correspondence to: bediako@berkeley.edu

## Abstract

Tailoring electron transfer dynamics across solid–liquid interfaces is fundamental to the interconversion of electrical and chemical energy. Stacking atomically thin layers with a very small azimuthal misorientation to produce moiré superlattices enables the controlled engineering of electronic band structures and the formation of extremely flat electronic bands. Here, we report a strong twist angle dependence of heterogeneous charge transfer kinetics at twisted bilayer graphene electrodes with the greatest enhancement observed near the 'magic angle' (~1.1º). This effect is driven by the angle-dependent tuning of moiré-derived flat bands that modulate electron transfer processes with the solution-phase redox couple. Combined experimental and computational analysis reveals that the variation in electrochemical activity with moiré angle is controlled by atomic reconstruction of the moiré superlattice at twist angles <2º, and topological defect AA stacking regions produce a large anomalous local electrochemical enhancement that cannot be accounted for by the elevated local density of states alone. Our results introduce moiré flat band materials as a distinctively tunable paradigm for mediating electrochemical transformations.

**Introduction**

In electrochemical reactions, the flow of charge from electrode to electrolyte drives chemical transformations at the interface, and such interfacial electron transfer reactions underpin key technologies that interconvert electrical and chemical energy.[1–5] Traditionally, controlled doping or compositional modification of materials has been among the primary knobs for altering the electronic structure of surfaces and tuning their physicochemical properties.[2,3] Advances in the development of atomically thin, so-called two-dimensional (2D) materials, have provided new avenues for controlling interfacial charge transfer and surface reactivity.[6,7] Unlike bulk 3D materials in which the electrochemically active surface represents only some fraction of the total material, individual layers of these low-dimensional systems are 'all surface'. As a result, their electronic properties are exceptionally manipulable by structural modifications and other external electromagnetic perturbations. Accordingly, in addition to compositional/chemical doping,[7,8] other degrees of freedom emerge for modulating heterogeneous charge transfer at 2D layers including defect engineering,[9,10] strain tuning,[11,12] and field modulation by electrostatic gates.[13] Specifically, defects and edge sites present at 2D surfaces are recognized as reactive hot spots in interfacial charge transfer processes owing to their modified electronic properties.[7] However, the small areal fraction of edge sites[14] constrains the maximized utilization of the whole surface. In addition, it is generally challenging to achieve rational control of atomic defects[10] and an optimized selectivity[15] of the defect-induced enhancement.

2D atomic layers are also the building blocks for the design of artificial van der Waals (vdW) materials through the deterministic assembly of multiple layers.[16] These vdW heterostructures possess weak interlayer interactions that allow arbitrary azimuthal orientations ('twist angles') between the 2D lattices to be independently controlled, introducing a unique degree of freedom for modulating the electronic properties of 2D materials.[17,18] Recent work has discovered an extraordinarily wide and growing collection of exotic physics in twisted 2D layers.[19–23] For example, twisting a graphene bilayer from a Bernal (AB) stacking configuration to a 'magic angle' (MA) of ~1.1° results in the formation of electronic bands with very weak dispersion in momentum space (so-called "flat" bands).[24,25] Recent work has shown that these flat bands produce an array of intertwined correlated electronic phases, including correlated insulating states,[20] unconventional superconductivity,[19,21,23] and orbital magnetism.[22,23] Analogous architectures comprised of other 2D transition metal compounds have also recently been found to display flat bands and associated correlated electron behavior,[26] revealing this concept as a general approach for precisely tailoring the electronic structures of 2D solids. In twisted 2D materials such as these, a form of "topological" defects is formed at the intersection of soliton walls between different stacking domains.[27] Apart from hosting low-temperature correlated electron physics, these topological defects are conceptually analogous to structural atomic defects that generate localized states that can impact interfacial charge transfer chemistry. While a handful of studies interrogating the effects of interlayer twist on chemical reactivity have been reported,[28–30] these experiments have been restricted to large twist angles (≥ 7º) well outside the flat band and 'magic angle' regime. The possibility of exploiting this concept of moiré flat-bands and topological defects to modulate interfacial charge transfer rates has therefore remained unexplored. Here, we probe the kinetics of a heterogenous electron transfer reaction at well-defined twisted bilayer graphene (TBG) surfaces and show that the kinetics can be strongly tailored by the interlayer moiré twist angle, $\theta_m$, at small angles (0.22º ≤ $\theta_m$ ≤ 5º).

**Results and Discussion**

In TBG, the azimuthal misorientation between two homologous layers creates an in-plane quasiperiodic modulation in the crystallographic registry, revealed as a moiré superlattice pattern that displays alternating regions of AA, AB/BA, and saddle-point SP stacking configurations (Fig. 1a).[31–33] The periodicity of the moiré supercells, also known as the moiré wavelength, $\lambda_m$, and the size of the mini-Brillouin zone of the moiré superlattice (Fig. 1b) are both highly dependent on $\theta_m$. Hybridization between adjacent Dirac cones results in an electronic band structure that is strongly modified by the twist angle[20,24] as shown in Fig. 1c for MA-TBG ($\theta_m$ = 1.1°). At this $\theta_m$, the bands around 0 eV become remarkably flat, creating a massively enhanced electronic density of states (DOS) concentrated within a narrow energy range of ~20 meV (Fig. 1d).[20,24] At larger $\theta_m$, the bands become more dispersive, with increasing bandwidths and higher energies associated with the locations of the van Hove singularities (vHS). As we now show, these moiré flat band-originated DOS enhancements at vHS play a critical role in governing the heterogenous electron transfer rate at a graphene–electrolyte interface, making $\theta_m$ a descriptor of the interfacial reaction kinetics. We also demonstrate that spatially localized moiré flat bands allow the reaction kinetics to be precisely tuned by the topological defect (AA stacking) density, which is modulated as a function of $\theta_m$.

We fabricated hexagonal boron nitride (hBN)-supported TBG samples with controlled interlayer twist angles in the range of 0.22° – 5.1° (see Methods section for details). Figure 1e shows an optical micrograph of a TBG/hBN sample in contact with a graphite flake as the electrical contact. The pristine surfaces of these graphene moiré superlattice samples are characterized with atomic force microscopy, AFM, and scanning tunneling microscopy (STM). Fig. 1f shows a typical constant-current STM image of an exemplary TBG/hBN sample displaying a clear moiré superlattice pattern as observed previously.[32] For ultraclean TBG surfaces such as these, the strong electronic localization of flat bands in real space makes it possible to visualize the AA sites as elevated spots, while AB/BA and SP sites appear lower in STM topography.[32,32,34] The twist angle ($\theta_m$ = 1.15 ± 0.05°) and uniaxial heterostrain ($\varepsilon$ = 0.21 ± 0.12%)—produced by stretching one graphene layer relative to the other—are measured by geometric analysis of the local moiré superlattice patterns obtained from STM maps that are collected from multiple arbitrary locations in each sample. This random sampling approach is used to ensure reasonably uniform local twists throughout a TBG sample prior to electrochemical measurements.

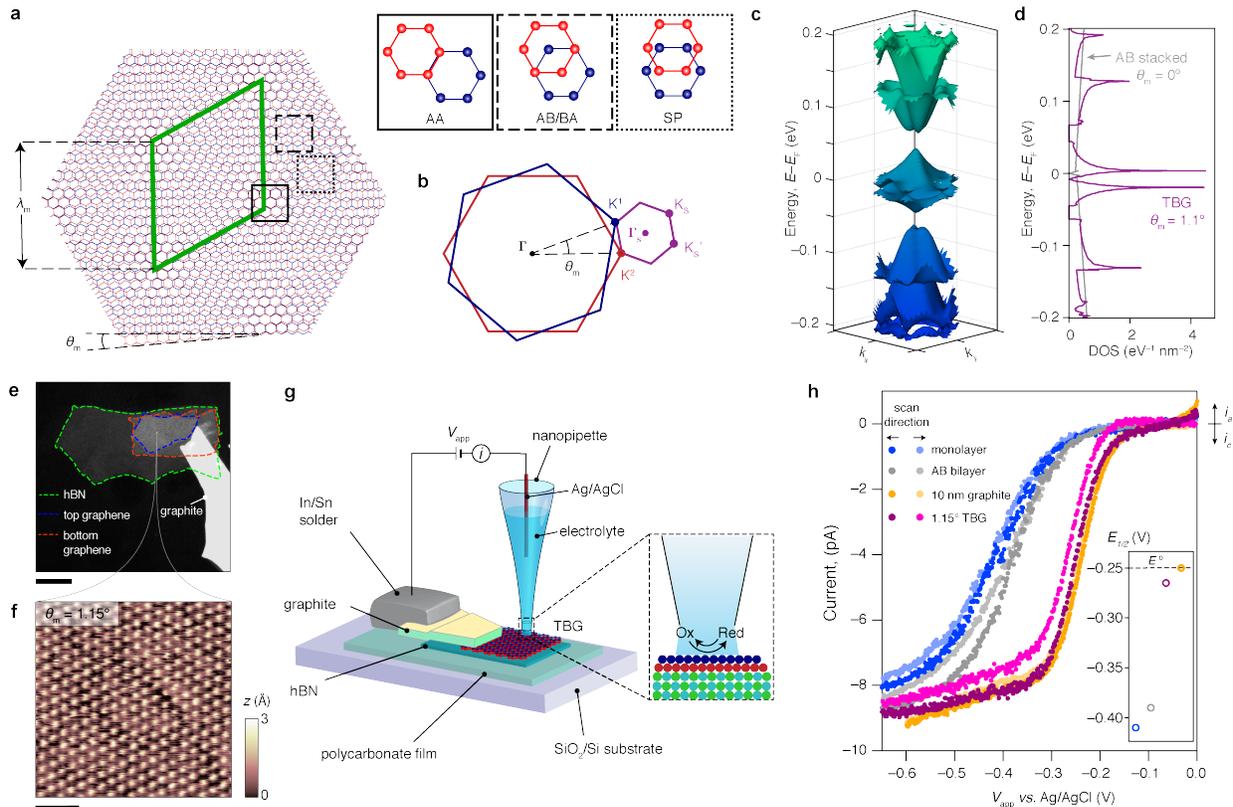

**Fig. 1. (a)** Schematic of a moiré pattern in TBG and the three stacking configurations (AA, AB/BA, and SP). The moiré wavelength $\lambda_m$ is determined by the interlayer twist angle $\theta_m$ by $\lambda_m = a/[2\sin(\theta_m/2)]$, where $a$ = 2.46 Å is the lattice constant of graphene. **(b)** The mini-Brillouin zone of a TBG superlattice generated from the difference between two wavevectors ($K^1$ and $K^2$). **(c–d)** Calculated moiré band **(c)** of 1.1° TBG and corresponding DOS **(d)**. $E_F$: fermi energy. **(e)** Optical image of a TBG/hBN heterostructure connected to a graphite contact. The dashed lines highlight the two graphene monolayers rotated by $\theta_m$ and the underlying hBN flake. Scale bar: 10 μm. **(f)** Constant-current STM image of 1.15° TBG acquired from the region marked with a yellow dot in **e**. The image was taken with a 0.1 V bias and a 100 pA set point. Scale bar: 50 nm. **(g)** Schematic of local voltammetric measurement at a TBG surface in an SECCM setup equipped with a single channel nanopipette probe (diameter ~100 nm). The electrochemical bias is applied to the TBG via the In/Sn and graphite contact with respect to the quasi-reference counter electrode (Ag/AgCl) inside the nanopipette filled with electrolyte. **(h)** Representative steady-state voltammograms of 2 mM $Ru(NH_3)_6^{3+}$ in 0.1 M KCl solution obtained at separate, freshly prepared graphene monolayer (blue), Bernal stacked bilayer (grey), 10-nm thick graphite (orange), and 1.15° TBG (purple). Scan rate: 100 mV s$^{-1}$. The inset shows the half-wave potentials of each CV compared to $E^o$.

Scanning electrochemical cell microscopy (SECCM) with *ca.* 100 nm spatial resolution allows us to obtain electrochemical measurements exclusively at the basal plane of TBG (Fig. 1g). A quartz nanopipette (diameter < 500 nm) filled with 2 mM hexaammineruthenium(III) chloride and 100 mM aqueous potassium chloride is used to make meniscus contact with the sample surface, creating a confined electrochemical cell in which localized voltammetry is performed at a series of locations.[14,15,35,36] Typical steady-state voltammograms of Ru(NH$_3$)$_6^{3+}$ reduction obtained at the basal planes of hBN-supported monolayer graphene, Bernal (AB) stacked bilayer graphene, 10 nm thick graphite (~30 layers), and 1.15° TBG are shown in Fig. 1h. The anodic shift of the half-wave potential, $E_{½}$ (potential at $i = i_\infty/2$, where $i_\infty$ is the diffusion-limited current plateau), with increasing graphene thickness is consistent with the electroreduction of Ru(NH$_3$)$_6^{3+}$ proceeding more rapidly with augmented DOS from increasing numbers of layers.[14] Remarkably, the 1.15° TBG shows a massive enhancement in reaction kinetics, approaching that of graphite notwithstanding its consisting of only two graphene layers, and exhibiting nearly electrochemically reversible behavior at the basal plane of this atomically thin electrode.

Interpreting the heterogenous electron transfer kinetics at these TBG surfaces mandates the consideration of some physicochemical properties that are integral to electrochemical interfaces of low-dimensional materials. Applying an electric potential ($V_{app}$) across a solid–solution interface results in the formation of an electrical double layer adjacent to the solid surface, which serves to screen the excess charge. At low dimensional materials like graphene, the charged ions of the double layer also act cumulatively as an effective electrostatic 'gate' that shifts the Fermi level, $\epsilon_F$, relative to the band edges of the material *via* dynamic electron/hole doping. Consequently, a significant fraction of $V_{app}$ is consumed as a change of the chemical potential ($V_q$), while the remainder is confined in the electric double layer ($V_{dl}$). The contribution from $V_q$ is modeled by introducing an additional capacitance, the so-called quantum capacitance ($C_q$),[37,38] in series with the double layer capacitance ($C_{dl}$) as shown in Fig. 2a. $C_q$ depends on the electronic structure of the material and is a function of the chemical potential. To evaluate $C_q$ in TBG, we computed electronic band structures for TBG over a range of $\theta_m$ (see Methods Section) and determined theoretical $C_q$ values as a function of $V_q$ in each case. Figure 2b shows that higher values of $C_q$ near the charge neutrality point are observed in TBG for $1° < \theta_m < 2°$, suggesting that a lower fraction of $V_{app}$ would be consumed on $C_q$ in TBG around these values of $\theta_m$ due to an enhanced DOS near $\epsilon_F$. To directly evaluate the effect of this $\theta_m$-tunable quantum capacitance on $V_{dl}$, Fig. 2c shows how $V_{dl}/V_{app}$ evolves as a function of $V_{app}$ as $\theta_m$ is tuned. Higher values of $V_{dl}/V_{app}$ are also observed near charge neutrality for $1° < \theta_m < 2°$, consistent with the angle dependence of $C_q$ in Fig. 2b. These results show how moiré-derived flat bands can promote bulk metal-like behavior (increased $C_q$ and higher $V_{dl}/V_{app}$) in low-dimensional electrodes.

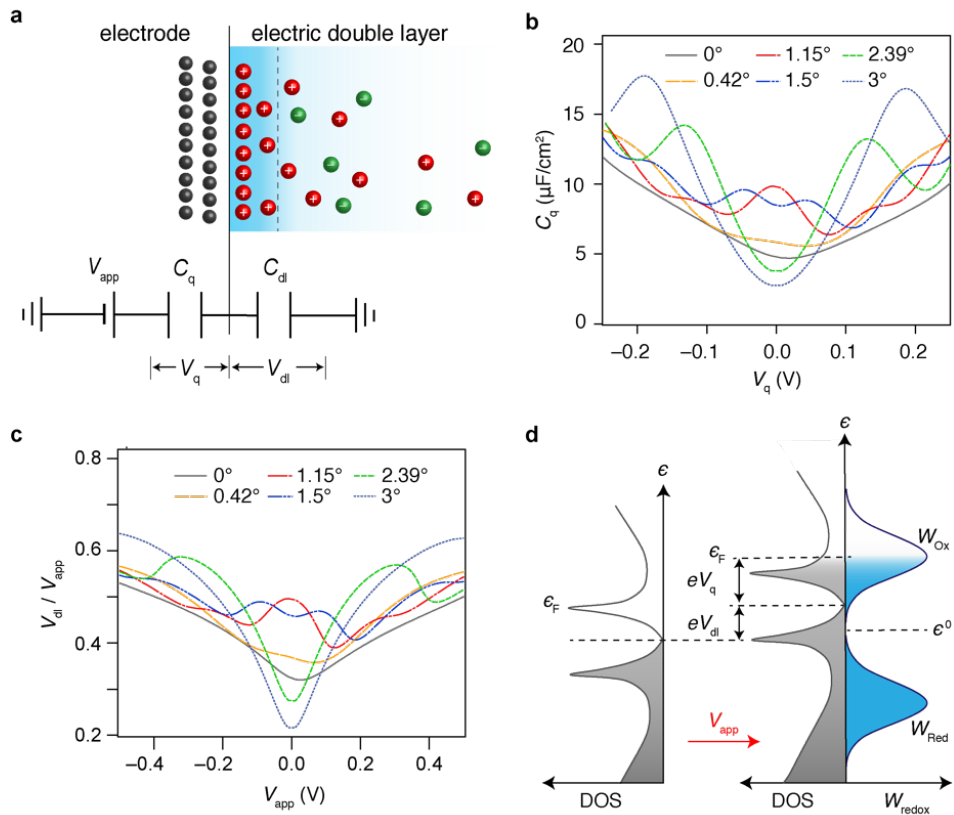

**Fig. 2. (a)** Schematic of the electrode-electrolyte interface represented as the double-layer capacitance ($C_{dl}$) in series with the chemical quantum capacitance ($C_q$). **(b)** Calculated $C_q$ as a function of the chemical potential ($V_q$) for various twist angles. **(c)** Calculated potential drop on double layer ($V_{dl}$) as a function of the applied potential ($V_{app}$) for various twist angles. $V_q$ and $V_{app}$ are relative to the charge neutrality potential. **(d)** Energy diagram depicting that the charge transfer rate is determined by the overlap of electronic states between the electrode (*DOS*) and the probability distributions representing the solution-phase redox molecules ($W_{ox}$ and $W_{red}$). Upon applying $V_{app}$, the Fermi level is shifted by $eV_q$ relative to the band edge, while the band structure is shifted $eV_{dl}$ with respect to the electronic states of the molecule.

To consider the impact of partitioning $V_{app}$ into $V_q$ and $V_{dl}$ in the electrochemistry of TBG, we turn to the Gerischer–Marcus [39,40] or Marcus–Hush–Chidsey (MHC) formalism with the consideration of electrode DOS,[41,42] which describes the heterogeneous electron transfer rate constant, $k_{red}$ (for an electroreduction reaction) in the weak coupling (outer-sphere) limit. In this framework, $k_{red}$ is governed by the distribution and overlap of the electronic energy states, and electron transfer takes place from any occupied state to any unoccupied receiving state with a matching energy. Ultimately, $k_{red}$ is the integral of the overlap between the electronic states of the electrode and those of the solution-phase molecule at all energies. When we include considerations brought about by a finite $C_q$, $k_{red}$ can be modeled as:

$$k_{red} = \nu \int_{-\infty}^{+\infty} D(\epsilon - eV_{dl}) f(\epsilon - eV_{app}) W_{Ox}(\epsilon) d\epsilon \tag{1}$$

$$W_{Ox} = (4\pi\lambda k_B T)^{-1/2} e^{\frac{-(\epsilon - \epsilon^0 - \lambda)^2}{4\lambda k_B T}} \tag{2}$$

where $\nu$ is the integral pre-factor that considers the strength of the electronic interactions between the reactant and electrode, $D(\epsilon)$ is the DOS of the electrode, $f(\epsilon)$ is the Fermi-Dirac function, $W_{Ox}(\epsilon)$ is the normalized probability distribution representing the electronic states of the reactant, $\lambda$ is the reorganization energy, $k_B$ is the Boltzmann constant, and $\epsilon^0$ is the energy corresponding to the standard potential, $E^0$. In Fig. 2d, we consider the result of applying a potential bias $V_{app}$ across the solid–electrolyte interface. In this case, $\epsilon_F$ is shifted by $eV_q$ relative to the band edge, while $eV_{dl}$ shifts the entire band relative to the energy of the redox molecule. This means that ultimately, the energetic alignment of electron donor–acceptor states is affected by the relative contributions from $V_{dl}$ and $V_q$, and $k_{red}$ is governed by $D(\epsilon)$ near $\epsilon_F$. Since $D(\epsilon)$ is controlled by $\theta_m$, this analysis predicts that $k_{red}$ should display a dependence on $\theta_m$.

Figure 3 examines this expectation. TBG samples with a range of $\theta_m$ (exemplary STM images are shown in Fig. 3a) are used to measure steady-state voltammograms of $Ru(NH_3)_6^{3+}$ reduction. We note that measured values of $\varepsilon$ are <1% in all our samples and therefore contribute minimally to perturbation of the interfacial charge transfer kinetics. On the basis of the established model for voltammetric SECCM experiments,[14] we performed finite element simulations of the characteristic voltammetric responses to extract the standard rate constant, $k^o$, from the experimental cyclic voltammetry (CV) data. This kinetic parameter $k^o$ denotes the intrinsic electron transfer rate defined in the Butler–Volmer (BV) formulation:[40]

$$k_{red}^{BV} = k^o e^{-\alpha \frac{F}{RT}(V_{app} - E^0)} \tag{3}$$

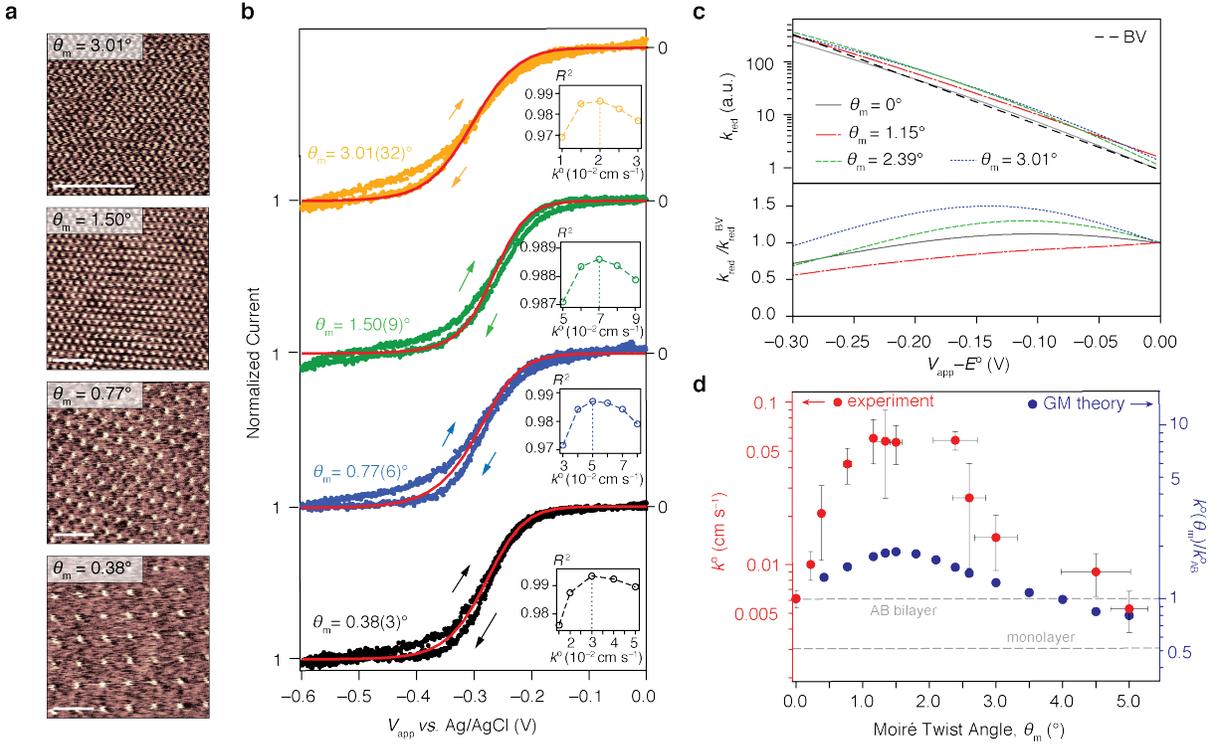

**Fig. 3. (a)** Constant current STM images of TBG with $\theta_m$ = 0.38°, 0.77°, 1.5°, and 3°. Scale bars: 50 nm. **(b)** Representative steady-state voltammograms of 2 mM Ru(NH$_3$)$_6^{3+}$ in 0.1 M KCl solution obtained at TBG with $\theta_m$ = 0.38°, 0.77°, 1.5°, and 3°. The current *vs.* potential curves were normalized to the mass transfer limiting current. The simulated voltammograms are depicted by solid red lines and the insets display the $R^2$ values from comparisons of experimental voltammograms to simulated curves using different rate constants. Scan rate =100 mV s$^{-1}$. **(c)** Top: $k_{red}$ *vs.* $V_{app}$ dependence for different $\theta_m$ calculated with Gerischer–Marcus (GM) framework compared to the trend of the BV model with a constant pre-factor (for clarity, the BV trace is shown only for the case where $\theta_m$ = 0°). Bottom: potential-dependent pre-factor, $A(V_{app}) = k_{red}/k_{red}^{BV}$. **(d)** Standard rate constants ($k^o$) extracted from the experimental voltammograms as a function of twist angle compared to the theoretical values. Each red dot denotes the mean value of the $k^o$ values obtained from a series of measurements at one sample. The horizontal and vertical error bars represent the standard deviations of $\theta_m$ and $k^o$, respectively. The dashed lines show the experimental $k^o$ values of monolayer and AB bilayer for comparison.

where $\alpha$ is the transfer coefficient and $F$ is the Faraday constant. In our analysis, we consider two key effects that arise from the effects of variable $C_q(V_q)$ with $\theta_m$ that were discussed in Fig. 2.

First, the change of electric double-layer structure as a consequence of $C_q$ may also be expected to exert influence on the mass transport of charged molecules, which would in turn affect the voltammetric response.[43,44] We therefore explicitly account for diffuse double layer effects involving the migration of charged species in the diffuse layer. Second, we introduce a potential dependent pre-factor $A(V_{app})$ to the BV equations to account for the possibility of changing DOS with $V_{app}$. We determine $A(V_{app})$ by integrating the convolution of $D(\epsilon)$, $f(\epsilon)$, and $W_{Ox}(\epsilon)$ over all energies to obtain theoretical $k_{red}$ values as a function of $V_{app}$ for each $\theta_m$ (Fig. 3c, top). We then compare these $k_{red}$ values to those derived from the BV model with a constant pre-factor, such that $A(V_{app}) = k_{red}/k_{red}^{BV}$ (Fig. 3c, bottom). Thus, the observed slight variations of $A(V_{app})$ for each $\theta_m$ originate from the disparate potential dependences of $k_{red}$ by virtue of changing $D(\epsilon)$ and $f(\epsilon)$. Figure 3c shows that the theoretical $k_{red}$ values that are derived from the Gerischer–Marcus model can be reasonably fit by the BV model with a constant pre-factor for small overpotentials ($V_{app} - E^0 < 0.2$ V), but deviations between these two cases become pronounced at larger overpotentials ($V_{app} - E^0 > 0.2$ V).[45] Due to the mass-transport limitations in the experiment, we only extracted kinetic information at low overpotentials (< 0.2 V). Nevertheless, we still apply this small correction for the potential dependence of the apparent rate constant in BV equations.

Figure 3d shows the resultant dependence of $k^o$ on $\theta_m$ extracted from the experimental voltammograms (red markers). As implied by the voltammograms in Fig. 1h, we find the $k^o$ extracted from TBG for 1° < $\theta_m$ < 2° is strongly enhanced over that of Bernal (AB)-stacked bilayer graphene ($\theta_m$ = 0°). By comparison to Bernal bilayer graphene, there is minimal enhancement of $k^o$ for $\theta_m$ > 4°, and for $\theta_m$ < 1°, $k^o$ decreases monotonically with $\theta_m$. We attribute the enhancement of $k^o$ between $\theta_m$ = 1° and $\theta_m$ = 2° to the elevated DOS arising from the flattened bands that are energetically matched with the formal potential of $Ru(NH_3)_6^{3+/2+}$. Our assertion on the criticality of energy matching with the flat bands in dictating the enhanced electron transfer kinetics at TBG was tested by measurements of the electrochemical kinetics of ferrocenemethanol (FcMeOH) and tris(1,10-phenanthroline)cobalt(II), $Co(Phen)_3^{2+}$, which possess formal potentials that are poorly aligned with the charge neutrality point of graphene and, for that matter, the flat bands of small angle ($\theta_m$ < 2°) TBG. For $FcMeOH^{+/0}$, inherently facile kinetics leads to nearly reversible voltammetric responses of FcMeOH[14,36] at all $\theta_m$. However, the intrinsically slow kinetics of $Co(Phen)_3^{3+/2+}$ allowed us to extract rate constants from the voltammograms obtained at various $\theta_m$, revealing a considerably weaker dependence of $k^o$ on $\theta_m$ compared with that of $Ru(NH_3)_6^{3+/2+}$. These results precisely match our expectation for energetically misaligned donor–acceptor electronic states involving the flat bands; the electrochemical kinetics are not substantially dependent on the electronic flat band.

Figure 3d also shows the theoretical dependence of $k^o$ for $Ru(NH_3)_6^{3+/2+}$ on $\theta_m$ (blue markers) calculated from the Gerischer–Marcus model (equations 1 and 2), manifesting a qualitatively consistent trend. The maximum enhancement is predicted between $\theta_m$ = 1.2° and $\theta_m$ = 1.6°, where the optimal combination of DOS and energy match is achieved after incorporating the effects of quantum capacitance. For $\theta_m > 2°$, the position of vHS become further apart from $E^0$, the magnitude of DOS gradually decays, and $k^o$ diminishes as a result. While qualitatively descriptive of the experimental data, interestingly, these calculations greatly underestimate the enhancement due to flattened bands in the range $0° < \theta_m < 3°$. To explain this substantial disparity between the Gerischer–Marcus model and our experimental results (the anomalous enhancement of $k^o$ at low twist angles) as well as the observed monotonic decrease in $k^o$ with decreasing $\theta_m$ for $\theta_m < 1°$, we evaluated the experimental and theoretical real-space variations in $k^o$ throughout the moiré superlattice.

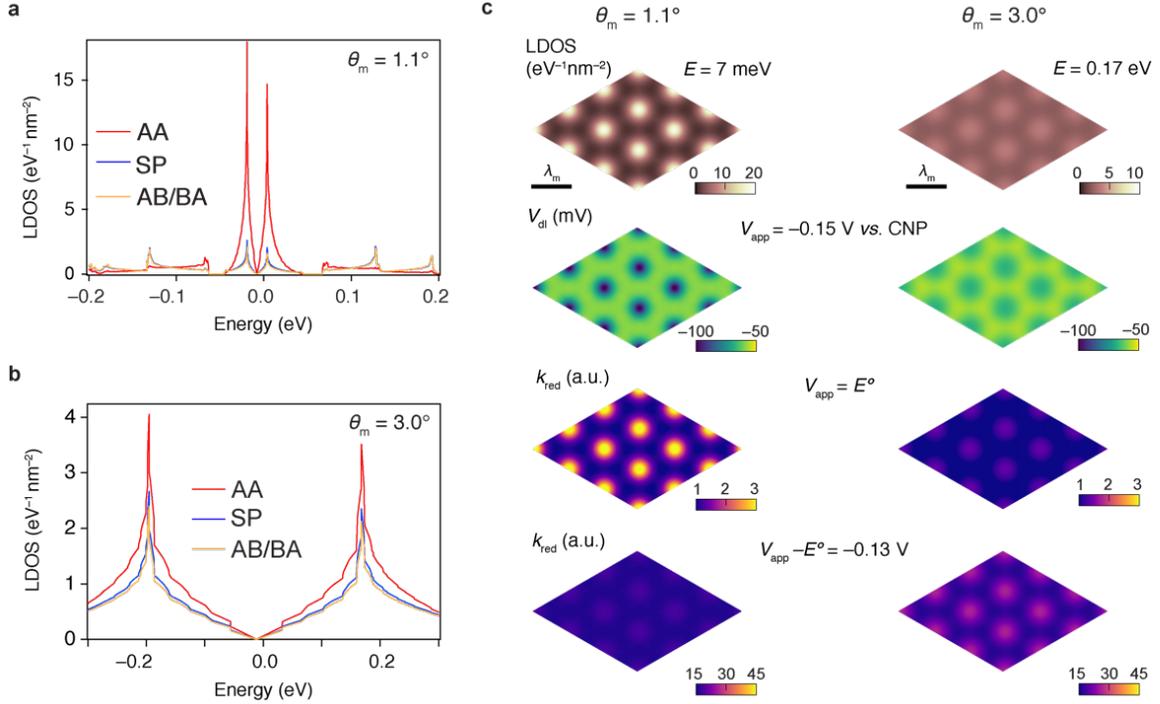

**Fig. 4. (a–b)** Calculated LDOS *vs.* energy at AA, AB/BA, and SP in TBG of $\theta_m = 1.1°$ (**a**) and 3° (**b**). **(c)** Calculated real space maps of LDOS, $V_{dl}$, and $k_{red}$ for TBG of $\theta_m = 1.1°$ (left column) and 3° (right column). The LDOS are mapped at the energy corresponding to the vHS of each angle. The $V_{dl}$ values are calculated at $V_{app} = -0.15$ V *vs.* charge neutrality potential (CNP). Maps of $k_{red}$ are shown at lower overpotential (third row) and higher overpotential (fourth row). The scale bars represent the length of $\lambda_m$ (12.8 nm for $\theta_m = 1.1°$, and 4.7 nm for $\theta_m = 3°$).

The electronic flat band formed in TBG moiré superlattices is strongly localized in real space. The local DOS (LDOS) *vs.* energy at the three stacking regions (AA, AB/BA, and SP) in MA-TBG ($\theta_m = 1.1°$) is shown in Fig. 4a. These calculations show how the prominent enhancement of DOS near charge neutrality is strongly localized at AA sites with much lower DOS found at AB/BA or SP stacking sites, in agreement with previous studies.[32,34] In contrast, Fig. 4b shows a very weak real-space partitioning of the overall DOS in TBG with $\theta_m = 3°$ where flat bands are largely absent. Instead, in this regime the LDOS is generally insensitive to the local stacking order. These localization effects can be visualized in the real space LDOS maps shown in Fig. 4c. Importantly, as a consequence of DOS localization, the interfacial properties of moiré superlattice electrodes are also strongly modulated in real space. Specifically, Fig. 4c also shows that the potential drop across the double layer ($V_{dl}$) should be nearly uniform in real space over the moiré structure for $\theta_m = 3°$, whereas $V_{dl}$ would be concentrated at AA sites for $\theta_m = 1.1°$ due to higher LDOS and $C_q$. Similarly, these calculations predict an enhancement of the electroreduction rate constant, $k_{red}$ for Ru(NH$_3$)$_6^{3+/2+}$ at AA sites near zero overpotential ($V_{app} = E°$). The massive LDOS at the matched energy results in significant augmentation of the integral between $D(\epsilon)$, $f(\epsilon)$, and $W_{Ox}(\epsilon)$ described in equations 1 and 2. As the overpotential increases, $\epsilon_F$ shifts over the flat band due to the charging of $C_q$. The enhancement effects from the flat band are diminished at high overpotentials ($V_{app} - E^0 = -0.13$ V), leading to a uniform $k_{red}$ distribution across the moiré unit cell. For $\theta_m = 3°$, the calculated maps of $k_{red}$ show minimal spatial variation over all potentials, as expected. Theoretical maps at $\theta_m < 1°$ resemble those of $\theta_m = 1.1°$, showing strong localization of LDOS, $V_{dl}$, and $k°$ at AA sites only scaled by the change of $\lambda_m$. To compare these theoretical results to the experiment, we measured the local stacking order-dependent electrokinetics over the moiré superlattice sites.

Even though experimentally, local electron transfer kinetics at isolated AA stacking regions are not directly measurable owing to the spatial resolution limit (*ca*. 100 nm) in our SECCM setup, we can reliably extract local $k°$ values for AA stacking regions ($k°_{AA}$) from the ensemble $k°$ measured in the meniscus covering multiple moiré unit cells. To do so, we take advantage of the independently measurable rate constant for AB stacking ($k°_{AB}$), which is determined from standalone measurements at AB bilayers (*i.e.*, $\theta_m = 0°$) (Fig. 1h and Fig. 3d). Considering the ensemble current as arising from the sum of local currents weighted by the area fraction, $A$, of each domain (AA, AB/BA and SP), the ensemble rate constant, $k°$, may be written as:

$$k^0 = A_{AA}k^0_{AA} + A_{AB}k^0_{AB} + A_{SP}k^0_{SP} \tag{4}$$

Where $k^0_{AA}$, $k^0_{AB}$, and $k^0_{SP}$ denote the respective local rate constants at AA, AB/BA, and SP domains. To simplify the analysis, we assume $k°_{AB} \cong k°_{SP}$ —a reasonable approximation given the very similar LDOS calculated (Figs. 4a–c) and measured by STM (Fig. 3b).[32] These considerations reduce equation 4 to,

$$k^0 = A_{AA}k^0_{AA} + (1 - A_{AA})k^0_{AB} \qquad (5)$$

To determine area fraction values for equation 5, we have previously employed four-dimensional scanning transmission electron microscopy, 4D-STEM[33], to reveal that $A_{AA}$ increases monotonically with $\theta_m$ until $\theta_m$ reaches ~2° due to a spontaneous lattice reconstruction process involving localized rotations,[31,33,46] after which area fractions approach values expected from a rigid moiré. From these established measurements we know $A_{AA}$, $A_{AB/BA}$, and $A_{SP}$ as a function of $\theta_m$. These insights allow us to extract experimental values for $k^o_{AA}$ as a function of $\theta_m$ in Fig. 5a. Interestingly, we find that $k^o_{AA}$ is insensitive to twist angle for $\theta_m < 1.5°$, showing a fairly constant value around 0.2 cm s$^{-1}$. This trend is remarkably consistent with our previous structural measurements[33] that unveiled the geometric effects of lattice reconstruction. This effect is depicted in Fig. 5b. At $\theta_m < 1.5°$, positive local rotations centered on the AA stacking sites, where flat band wave-functions are predominantly localized (Fig. 4c), cause the atomic geometry in these regions to be nearly insensitive to the global $\theta_m$, 'pinning' the local rotation to ~1.2°. On the other hand, localized negative rotation in AB/BA regions produces a local atomic geometry that closely matches Bernal stacked bilayer graphene.[33] Thus, our measurements show that for $\theta_m < 1.5°$, the primary effect that drives the $\theta_m$-dependent ensemble $k^o$ shown in Fig. 3d is the change in the moiré wavelength (and attendant AA stacking area fraction), whereas the local electrochemical activities remain effectively constant (plateau in $k^o_{AA}$ for $\theta_m < 1.5°$ in Fig. 5a). This convergence of local electrochemical behavior is also manifest in plots of $k^o_{AA}$ as a function of the net local AA rotation after lattice relaxation, $\theta_T^{AA}$ (inset of Fig. 5a). At larger $\theta_m$, the enhancement from flat bands and the atomic reconstruction process both diminish, causing $k^o_{AA}$ to eventually become similar to $k^o_{AB}$ for $\theta_m \geq 4°$. As a result, $k^o_{AA}$ exhibits a monotonic downward trend with the increase of the local and global moiré angle at AA sites (inset of Fig. 5a).

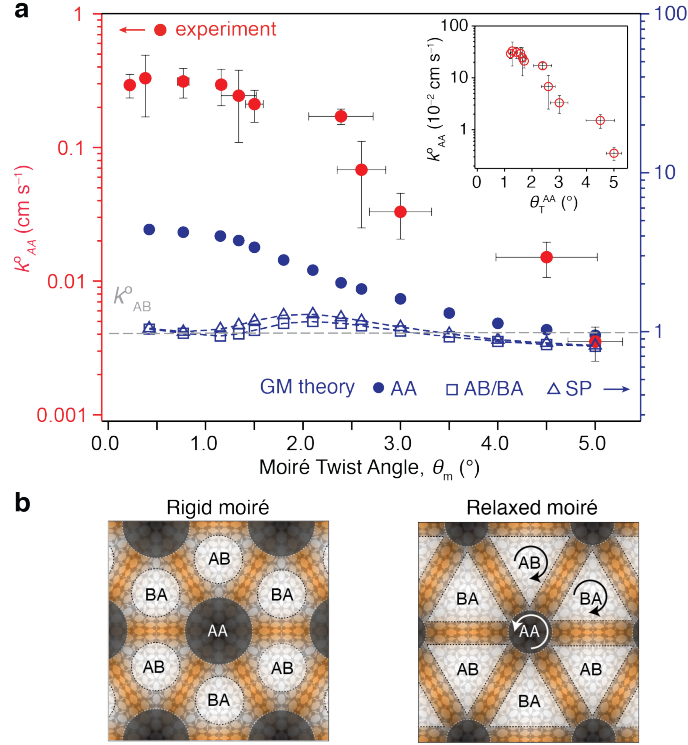

**Fig. 5.** (a) Experimental local $k^o$ at AA sites as a function of $\theta_m$ extracted from Fig. 3d compared to the theoretically predicted (Gerischer–Marcus, GM) values of $k^o$ at AA, AB/BA, and SP stacking regions. The horizontal and vertical error bars represent the standard deviations of $\theta_m$ and $k^o_{AA}$, respectively. The dashed line shows the experimental $k^o$ values of AB bilayer represented as $k^o_{AB}$. The inset shows the local $k^o$ at AA sites as a function of the local net rotation angle of AA sites due to lattice relaxation, $\theta_T^{AA}$. (b) Schematic illustration of the lattice relaxation (atomic reconstruction) process, which modifies the area fraction and local net interlayer rotation angle of the various stacking configurations through localized rotational deformations of the moiré superlattice.

While this local $\theta_m$ dependence is qualitatively supported by the theoretical $k^0_{AA}$, $k^0_{AB}$, and $k^0_{SP}$ values that are plotted in parallel with the experimental values in Fig. 5, we find a large discrepancy between the experimental $k^o_{AA}$ values and those calculated using the Gerischer–Marcus model alone. Crucially, Fig. 5 reveals that the discrepancy between theory and experiment in overall $k^o$ (Fig. 3d) originates from this underestimation in $k^o_{AA}$ specifically (Fig. 5a). Further, this large, anomalous enhancement of the electron transfer rate (over that predicted from the Gerischer–Marcus theory) near and below the MA, which originates from the topological defects of the moiré with AA-localized flat bands, cannot be explained merely by the commensurability of energetic overlap between donor–acceptor states, nor by the enhanced DOS (Fig. 2d). Instead, this observation may be explained by modifications to the electronic coupling strength ($v$ in equation 1) with $\theta_m$. At present, our theoretical analyses in Fig. 3d and Fig. 5 do not consider a dependence of $v$ on $\theta_m$ via changing local stacking configurations and band flatness. However, it is reasonable to expect that $v$ should be strongly impacted by the localized flat band structure and this should then make a considerable contribution to the electron transfer rate according to equation 1. Indeed, the electronic transmission coefficient is increased at graphene zigzag edges in the non-adiabatic regime owing to localized electronic edge states at $\epsilon_F$.[47] By analogy, the highly localized flat bands in small angle TBG (Fig. 4c) would result in significantly larger real-space overlap of the electronic wave functions of the TBG and solution-phase redox complexes, inducing an additional augmentation of the measured rate. The effects of moiré reconstruction and pinning of the local rotation field near the MA (~1.2°)[33] would then keep this enhancement constant for $\theta_m < 2°$, precisely as observed in Fig. 5a. Additionally, it is conceivable that greater local potentials (higher local $V_{dl}$ in Fig. 4c) at AA sites could also modify the structure of the electric double layer and/or create a shorter electron tunneling distance between the electrode and redox species, through electroosmosis and/or stronger electrostatic interactions with the charged complexes, and thus a transition to the adiabatic regime.[40]

**Conclusions**

The intriguing interfacial electron transfer behavior of TBG demonstrated here suggests that in addition to low-temperature correlated electronic phases, moiré-derived flat bands in vdW architectures present a highly tunable material platform to systematically manipulate and fundamentally probe interfacial charge transfer and (electro)chemical transformations at well-defined surfaces. In lieu of introducing foreign dopants or structural defects that can be difficult to control, we have demonstrated the feasibility of activating the pristine graphene basal plane surface with topological defects that originate from the structural symmetry breaking in twisted bilayers.[27] The AA stacked sites in which the flat bands are strongly localized bear analogy to the structural defects that have been shown to produce enhanced DOS and promote interfacial charge transfer.[7,10,14,47] However, here we show, simply by the precise control of the interlayer twist angle, flat bands that are intrinsically localized to nanoscale domains in real space, may be exploited to control the rate constant for an outer-sphere charge transfer reaction. We anticipate that flat-band engineering with moiré architectures is a general strategy to tune the chemical reactivity of a variety of 2D surfaces. The wide array of 2D transition metal compounds[48] and the generality of the moiré superlattice approach for engineering flat bands in these materials as well[26] foreshadows this 'twistronic' topological defect and moiré flatband approach as a versatile new avenue for also tailoring (electro)catalytic (inner-sphere) reactions where interfacial reactivity is governed by the structure and dispersion of the electronic bands of the electrode near $\epsilon_F$, which determine the adsorption energetics of intermediates.[3,39]

**Methods**

**Chemicals and Materials.** Natural Kish graphite crystals were purchased from Graphene Supermarket. Si/SiO$_2$ wafers (0.5 mm thick with 285 nm SiO$_2$) and polydimethylsiloxane (PDMS) stamps were purchased from NOVA Electronic Materials, LLC and MTI Corporation, respectively. Sn/In alloy was purchased from Custom Thermoelectric, INC. Poly(bisphenol A carbonate), hexaammineruthenium(III) chloride (98%), ferrocenemethanol (97%), cobalt(II) chloride hexahydrate (98%), 1,10-phenanthroline (>99%), potassium chloride (>99%), and 4-Nitrobenzenediazonium tetrafluoroborate (97%) were purchased from Sigma-Aldrich and used as received. Sulfuric acid (ACS grade, 95% – 98%) was used as received from Lab Chem (Fisher Scientific). All aqueous electrolyte solutions were prepared with type I water (EMD Millipore, 18.2 MΩ cm resistivity). 2 mM solutions of tris(1,10-phenanthroline)cobalt(II) were prepared by dissolving 1:3 molar ratios of solid cobalt(II) chloride and 1,10-phenanthroline in water. In Ru(NH$_3$)$_6^{3+}$ and Co(phen)$_3^{2+}$ solutions, solid KCl was added as a supporting electrolyte at a concentration of 100 mM.

**Sample fabrication.** Graphene and boron nitride (hBN) flakes were mechanically exfoliated onto SiO$_2$(285 nm)/Si substrates from their bulk crystals using the well-established "scotch tape" method. Twisted bilayer graphene (TBG) samples were prepared using the "tear and stack" technique[20] on a temperature-controlled heating stage (Instec, Inc.) equipped with an optical microscope (Mitutoyo FS70) and a micromanipulator (MP-285, Sutter Instrument). A thin poly(bisphenol-A carbonate) (PC) film attached onto a polydimethylsiloxane (PDMS) stamp was used to pick up individual hBN flakes (10–20 nm thick) on the SiO$_2$/Si substrate. The picked-up hBN was aligned with half of a graphene monolayer, which was then torn along the edge of the hBN. The stage was rotated to a desired moiré twist angle ($\theta_m$), then the picked-up graphene was placed over the remaining portion of the graphene to form the TBG/hBN heterostructure. A thick graphite flake (10–100 nm thick) was then picked up to partially overlap with the graphene to form the immediate electrical contact to the TBG. The PC film was then delaminated from the PDMS stamp and placed onto a fresh SiO$_2$/Si chip. Sn/In contacts were made to the graphite contact using the micro-soldering method[49].

**STM and electron microscopy measurements.** STM measurements were carried out with a Park NX10 STM module (Park System) operated in ambient conditions (*i.e.*, room temperature, 1 atm pressure). Pt-Ir tips used for imaging were fabricated *via* electrochemical etching of 0.25 mm Pt-Ir wires in saturated $CaCl_2$ solutions. The transmission electron microscopy images of the nanopipettes were obtained with a JEOL 1200EX TEM operated at 100 keV. The ~1 mm portion of the pipette was attached to the grid (PELCO Hole Grids) such that the pipette tip is positioned in the center hole, and the rest of the pipette was cut off. Selected area electron diffraction (SAED) patterns were collected on an FEI Tecnai T20 S-TWIN TEM with a $LaB_6$ filament operated at 200 kV. SAED was used to resolve the twist angles for samples with $\theta_m > 5°$. To obtain the diffraction patterns, the fabricated TBG/hBN samples were transferred onto a holey silicon nitride membrane after electrochemical measurements.

**SECCM measurements.** The SECCM probes were single-channel nanopipettes pulled from quartz capillaries (ID = 0.7 mm, OD = 1 mm) using a laser puller (P-2000, Sutter Instrument), yielding orifices with sizes of 50 – 500 nm. The dimensions and geometries of the probes were determined by transmission electron microscopy (TEM). The nanopipettes were filled with the electrolyte solution containing redox species. Silver wires coated with AgCl were inserted into the channels of nanopipettes and used as quasi-counter/reference electrodes. The SECCM experiments were performed using Park NX10 SICM Module (Park Systems). The nanopipette was positioned over the area of interest aided by an optical microscope, then slowly approached the sample (at 100 nm/s) until a current > 2pA was detected due to the contact between the meniscus and the sample surface. A bias was held at a potential where the reaction takes place in a diffusion limited rate (–0.5 V *vs* Ag/AgCl for $Ru(NH_3)_6^{3+}$ and 0.5 V *vs* Ag/AgCl for $Co(Phen)_3^{2+}$) during the approach. The local cyclic voltammograms were subsequently recorded at various locations on the sample surface. For diazonium deposition on graphite, which was used to confirm that nanopipette sizes measured by TEM were representative of the meniscus contact area (as demonstrated previously [50,51]), deposition was performed using a nanopipette filled with 0.1 mM 4-Nitrobenzenediazonium tetrafluoroborate in 50 mM $H_2SO_4$ solution and with a constant bias of –0.2V *vs* Ag/AgCl for 20 seconds.

**Calculation of DOS.** The density of states (DOS) for TBG was calculated as a function of $\theta_m$ using the *ab-initio k·p* perturbation continuum model developed previously [52]. The model incorporates in-plane and out-of-plane relaxation effects as correction terms to intra-layer Dirac Hamiltonians and inter-layer coupling terms respectively[53]. The low energy electronic structure is based on a momentum expansion about the valley K point of the supercell Brillouin Zone (BZ) allowing a smooth dependence of bands on the twist angle, for $0.18° < \theta_m < 6°$. Previous work[52,54] shows that the *k·p* model exactly reproduces results of the more expensive *ab-initio* tight-binding model[55], and both are in good agreement with full density functional (DFT) theory calculations[56]. The energy range of integration for the DOS was fixed at ±0.7 eV around the charge neutrality point (CNP). For evaluation of the local DOS (LDOS), the normalized moiré supercell was divided into a 90 × 90 grid in real space and sampled over 36 k-points in the BZ. We kept the sublattice symmetry intact and assumed no extra screening of the inter-layer coupling constants. The fermi velocity, however, was increased by 20%, since DFT calculations of graphene without GW screening corrections are known to overestimate the experimental Fermi velocity by roughly 20%.[32] The DOS of $\theta_m = 0°$ was obtained from calculations of a single bilayer unit-cell with Bernal stacking, employing the same DFT-based tight-binding model as the twisted structures[55].

**Finite-element simulations.** The finite element simulations were performed using COMSOL Multiphysics v5.4 (COMSOL) to model the voltammetric responses. The "transport of diluted chemical species" module coupled with the "electrostatic" module was used to simulate the mass transport problems of redox species in the electrochemical cell defined by the nanopipette and the meniscus.

**Acknowledgments**

The authors acknowledge discussions with R. Kurchin. Research supported by the U.S. Department of Energy (DOE), Office of Science, Basic Energy Sciences (BES), under Award # DE-SC0021049 (experimental studies by Y.Y, K.Z., D.K.B.) and the Office of Naval Research under Award # N00014-18-S-F009 (computational work by H.P., M.B., and V.V.). S.C. acknowledges support from the NSF under grant # OIA-1921199. I.C. acknowledges support from a UCB Berkeley Fellowship. M.V.W. acknowledges support from an NSF GRFP award and UCB Chancellor's Fellowship. Confocal Raman spectroscopy was supported by a DURIP grant through the Office of Naval Research under Award # N00014-20-1-2599 (D.K.B.). D.K.B. acknowledges support from the Rose Hills Foundation through the Rose Hills Innovator Program. K.W. and T.T. acknowledge support from the Elemental Strategy Initiative conducted by the MEXT, Japan, Grant Number JPMXP0112101001, JSPS KAKENHI Grant Number JP20H00354 and the CREST (JPMJCR15F3), JST.

**Author contributions**

Y.Y. and D.K.B. conceived the study. Y.Y., K.Z., and A.L. performed the experiments. Y.Y. performed the COMSOL simulations. H.P., M.B., S.C., and V.V. carried out the theoretical calculations. I.C. performed the quantum capacitance calculations and STM image analysis. M.V.W. carried out the electron diffraction measurements. T.T. and K.W. provided the hBN crystals. Y.Y., K.Z., I.C., and D.K.B. analyzed the data. Y.Y. and D.K.B. wrote the manuscript.

**Correspondence and requests for materials** should be addressed to D.K.B. (e-mail: bediako@berkeley.edu).